\documentclass{article}
\usepackage{cite}
\usepackage{spconf,amsmath,graphicx}
\usepackage{algorithmic}
\usepackage{graphicx}
\usepackage{caption}
\usepackage{subfigure}
\usepackage{textcomp}
\usepackage{subfigure}
\usepackage{bm}

\title{Spatial Sparse subspace clustering for Compressive Spectral imaging}
\name{Jianchen Zhu$^{1,3}$, Tong Zhang$^{1,2}$, Shengjie Zhao\sthanks{Corresponding author: Shengjie Zhao }$^{1,2,3}$, Carlos Hinojosa$^4$, Zengli Liu$^5$, Gonzalo R. Arce$^6$}
\address{$^1$Key Laboratory of Embedded System and Service Computing, Ministry of Education \\ $^2$School of Software Engineering, $^3$School of Electronic and Information Engineering  \\ Tongji University, Shanghai, 201804, CHN \\ $^4$Department of System Engineering and Informatics \\ Universidad Industrial de Santander, Bucaramanga, 680002, COL \\ $^5$Institute of information engineering and automation Computer Center \\ Kunming University of Science and Technology, Kunming, 650500, CHN \\ $^6$Department of Electrical and Computer Engineering \\ University of Delaware, Newark, Delaware, 19716, USA \\ Email: zhujianchen@tongji.edu.cn, shengjiezhao@tongji.edu.cn.}
\begin{document}
%
\maketitle
\begin{abstract}
This paper aims at developing a clustering approach with spectral images directly from CASSI compressive measurements. The proposed clustering method first assumes that compressed measurements lie in the union of multiple low-dimensional subspaces. Therefore, sparse subspace clustering (SSC) is an unsupervised method that assigns compressed measurements to their respective subspaces. In addition, a 3D spatial regularizer is added into the SSC problem, thus taking full advantages of the spatial information contained in spectral images. The performance of the proposed spectral image clustering approach is improved by taking optimal CASSI measurements obtained when optimal coded apertures are used in CASSI system. Simulation with one real dataset illustrates the accuracy of the proposed spectral image clustering approach.
\end{abstract}
\section{Introduction}
\label{sec:intro}
Traditional spectral imaging (SI) techniques combine the 2D imaging and spectroscopy to sense spatial information across a multitude of wavelengths. The acquired 3D datacube can be viewed as spectral images, where two of the coordinates correspond to the spatial domain and the third one represents the spectral wavelengths. The disadvantage of these techniques is that the volume of datacube grows in proportion to the desired spatial or spectral resolution, and therefore exponentially growing the cost and time of data acquisition. Since the amount of radiation that each material reflects, scatters, absorbs, or emits depends on the wavelength, the spectral signature is valuable in many applications such as classification \cite{Crocco2017Audio}, target detection \cite{Nasrabadi2013Hyperspectral}, and spectral unmixing \cite{rajabi2015spectral} and so on.

With assuming that spectral signatures, which correspond to a land cover class, lie in the same low-dimensional subspace, spectral-based methods such as sparse subspace clustering (SSC) \cite{elhamifar2012sparse} has been proposed to build the adjacent matrix by  expressing each spectral pixel as a linear combination of all spectral signatures of the scene. Besides, SSC is realized by solving an $ l_1 $-minimization problem, whose solution is limited to be sparse in order to guarantee that the spectral signatures that correspond to those sparse coefficients belong to the same subspace. However, spectral image clustering is usually a challenging task due to the high-dimensional spectral data sets, which increase complexity and computational cost. Therefore, to mitigate these problems, it is necessary to reduce the dimensionality of spectral images. 

The SSC algorithm has been sucessfully applied to perform the spectral image clustering acquired using compressive spectral imaging (CSI) systems. Compressive spectral imaging systems require fewer compressed measuments than thoses obtained with traditional spectral imaging sensors. Our work aims at the clustring of the data acquired by a novel compressive imager, which is known as the spatial-spectral coded compressed spectral imager (3D-CASSI) system. The 3D-CASSI system first encodes spatial and spectral information of a scene using a 3D coded aperture and then the coded information is integrated along the spectral dimension. The 3D-CASSI system is different from the system in \cite{martin2016hyperspectral,cao2016computational,arguello2014colored,Yuehao2011Development,Arguello2013Higher} because that each spatial position of the acquired measurements contains the compressed information of a single coded spectral signature\cite{Xun2016Computational}.

Assuming that the compressed measurements lie in the union of multiple low-dimensional subspaces, this paper focuses on the unsupervised classification of every spectral pixel in the scene into one of the known classes from the given set of CASSI compressive measurements. The proposed approach is based on the SSC model where each spectral signature from its own subspace is represented as a sparse linear combination of all the spectral pixels, which ensure that the non-zero entires belong to the same class. Further, similar materials are represented as the neighboring pixels in a spectral image, which can help to extract more information from the data and can reduce the representation error by applying a smooth filter to the sparse matrix\cite{zhang2016spectral–spatial}.

The main contributions of this paper are twofold. First, the coded apertures used in the CASSI are realized by a greedy pursuit (GP) algorithm such that optimal compressed measurements are acquired, allowing the performance of spectral image clustering to be improved on comparison with the clustering performance obtained when traditional randomly coded pastures in CASSI are used \cite{wagadarikar2008single,Arguello2015Restricted}. Second, a compressive spectral image clustering approach is formulated, which is based on spectral pixel clustering directly from the compressed measurements.

\section{PROBLEM FORMULATION}
\label{sec:format}
\begin{figure*}
	\centering
	\includegraphics[width=14cm, height=5.5cm]{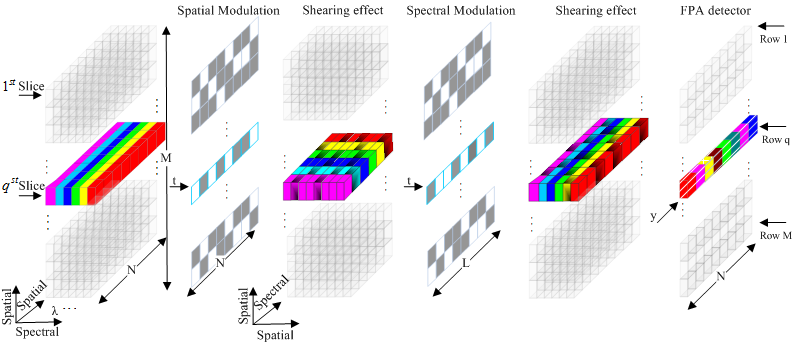}
	\caption{Illustration of the spatial-spectral optical flow in CASSI. The $ q^{th} $ slice of the datacube $ F $ with $ L=6 $ spectral components is coded by a row of the coded aperture $ t $ and sheared by the dispersive element. The detector acquires the intensity $ y $ by integrating the coded light.}
	\label{fig:figure2}
\end{figure*}

In 3D-CASSI system, as shown in Fig. 1, the voxels of the spectral scene is first modulated by using a 3D coded aperture. The coded spectral pixels are then integrated in the focal plane array detector (FPA) detector, along the spectral axis. Let $ T_{i,j,k}^{s} $ be the time-varying tridimensional coded aperture in its discrete form and $ F_{i,j,k} $ be the discretization of the source, where $ i $, $ j $ index the spatial coordinates, $ k $ the spectral component, and $ s $ the temporal component. The $ s^{th} $ discrete output on the FPA can be expressed as 
\begin{equation}\label{key}
{Y_{m,n}^{s}} = \sum\limits_{k = 0}^{L - 1} {T_{m,n,k}^{s}}{{F_{m,n,k}}}+w_{m,n}，
\end{equation}
where $ {Y_{m,n}^{s}} $ denotes the attained measurement at the $ (m,n)^{th} $ position on the detector at a specific snapshot $ s $ whose dimensions are $ M\times N $ and $ w_{m,n} $ is the white noise of the sensing system.  

Equation (2) can be rewritten in a linear matrix form as 
\begin{equation}\label{key}
y^{s}=H^{s}f+e,
\end{equation}
where $ y^{s}\in R^{MN} $ and $ F_{m,n,k}\in R^{MNL} $ are the vectorized representation of $ {Y_{m,n}^{s}} $ and $ {C_{m,n,k}^{s}} $, respectively, and $ H $ is the measurement matrix of the CASSI system, which is determined by the coded aperture pattern $ {T_{m,n,k}^{s}} $. The ensemble of $ S $ measurements can be expressed as 
\begin{equation}\label{key}
y^{S}=H^{S}f+e,
\end{equation}
where $ y^{S}=[(y^{0})^{T},...,(y^{S-1})^{T}] $ and $ H $ is the concatenation of matrices $ H^{s} $, $ s=0,...,S-1 $. Alternatively, the matrix of $ S $ coding pattern is defined as $ H=[H^{0},H^{1},...,H^{S-1} ]^{T} $ and $ f=[f_{0}^{T},...,f_{L-1}^{T}]^{T} $ is a $ L\times MN $ matrix whose columns are the spectral signatures $ f_{j} $ of the data cube. The ensemble of $ S $ measurements can be expressed as $ y^{S}=[(y^{0})^{T},...,(y^{S-1})^{T}]^{T} $ where $ y^{S} $ is a $ S\times MN  $ matrix. Notice in matrix $ y^{S} $ that each column value and each row value correspond to a compressed spectral signature and the compressed information (spectral response) of each pixels obtained at $ s^{th} $ snapshot, respectively. Then, the matrix $ y $ is convenient for SSC due to its structure, which makes easy to discriminate among compressed measurements.

\section{Proposed Algorithm for Coding Pattern Optimization}
\label{sec:pagestyle}
Inspired by curve-fitting techniques \cite{Liu2014Adaptive}, we utilize a smooth function to obtain the information from the given sets of neighboring spectral bands of interest, which leads to the preservation of the original signal structure.

Let $ (\left \{ \lambda _{1}^{S},\lambda _{2}^{S} \right \})=(\left \{ \lambda _{1}^{0},,\lambda _{2}^{0} \right \},...,\left \{ \lambda _{1}^{S-1},,\lambda _{2}^{S-1} \right \}) $ be the set selected to the matrix $ (H^{S})_{k}=\delta _{\left \lfloor \lambda _{1}^{S} \right \rfloor}\delta _{\left \lfloor \lambda _{2}^{S} \right \rfloor}h_{k}^{S} $, then the optimization problem can be expressed as  
\begin{equation}\label{key}
\begin{split}
&\underset{H,\lambda _{1}^{S},\lambda _{2}^{S},h^{s}}{min}f(H)=\left \| (H_{k}^{T})H_{k^{'}} \right \|_{F}^{2}+\left \| H^{s}(H^{s^{'}}) \right \|_{F}^{2} \\
&~~~s.t.~H\in \mathcal{C}_{L,S}, ~(H^{s})_{k}=\delta _{\left \lfloor \lambda _{1}^{s} \right \rfloor}\delta _{\left \lfloor \lambda _{2}^{s} \right \rfloor}h_{k}^{s} \\
&~~~~~~~\Delta (\lambda _{2}^{s}-\lambda _{1}^{s})=\Lambda-1, ~\det(H)\neq 0\\
\end{split}
\end{equation}
for $ k=0,...,L-1 $ and $ s=0,...,S-1 $.

The formulation given in (4) can be solved by applying the greedy pursuit (GP) as shown in Algorithm 1 because it can reduce computational complexity and improves computational efficiency.

\begin{table}[htb]
	\centering
	\label{}
	\begin{tabular}{p{2.9in}}
		\hline
		$\bm{ Algorithm }$ $ \bm{ 1 } $$ : $ Generate the optimal coding pattern  \\ \hline
		$\bm{ Input }$$ : $ $ L $, $ S $, $ \Lambda  $                             \\
		$\bm{ Output }$$ : $ $ H\in \left \{ 0,1 \right \}^{S\times L} $                            \\
		1: Initialize: $ (H^{0})_{k}\leftarrow \delta _{\left \lfloor \lambda _{1}/k \right \rfloor}\delta _{\left \lfloor k/\lambda _{2} \right \rfloor}h_{k}^{0}\sim Be(\frac{1}{2}) $                                 \\
		2: $\bm{ for }$ $ s_{it}\leftarrow 1 $ to $ S-1 $ $\bm{ do }$                                               \\
		3:~~~~~~$\bm{ for }$ $ k^{'}\leftarrow 0 $ to $ (L-\Lambda ) $ $\bm{ do }$                                              \\
		4:~~~~~~~~~~$ u_{k^{'}}\leftarrow \sum_{s^{'}=0}^{s_{it}}\sum_{k=k^{'}}^{k^{'}+\Lambda -1}(H^{s^{'}})_{k} $                                              \\
		5:~~~~~~$\bm{ end }$ $\bm{ for }$                                              \\
		6:~~~~~$ \lambda _{1}^{s_{it}},~\lambda _{2}^{s_{it}}\sim U_{ran}[argmin_{k^{'}}u_{k^{'}}]   $                        \\
		7:~~~~~~$ k^{'}k^{'}=0 $                      \\
		8:~~~~~~$\bm{ for }$ $ k^{'} \leftarrow \lambda _{1}^{s_{it}}~to~\lambda _{2}^{s_{it}} $                                              \\
		9:~~~~~~$ u_{k^{'}k^{'}}\leftarrow \sum_{s^{'}=0}^{s_{it}}\prod_{k=(k^{'}-1)}^{k^{'}}(H^{s^{'}})_{k} $, with $ h_{k^{'}}^{s_{it}}\leftarrow 0 $                                              \\ 
		10:~~~~~$ k^{'}k^{'}=k^{'}k^{'}+1 $                                              \\ 
		11:~~~~~$\bm{ end }$ $\bm{ for }$                                             \\
		12:~~~~~$\bm{ for }$ $ {k^{''}}\leftarrow 0 $ to $ \left \lfloor \frac{1}{2}\Lambda  \right \rfloor  $ $\bm{ do }$                                           \\ 
		13:~~~~~~~~~~$ \Gamma \sim U_{ran}^{'}[argmin_{k^{'}k^{'}}u_{k^{'}k^{'}}] $          \\
		14:~~~~~$ h_{\lambda_{l}^{s_{it}}}^{s_{it}}\sim Be(\frac{1}{2}) $, with $ \lambda _{l}^{s_{it}}\subset (\lambda _{1}^{s_{it}},\lambda _{2}^{s_{it}}) $                                                  \\
		15:~~~~~$\bm{ end }$ $\bm{ for } $                                           \\
		16:~~~~~$ (H^{s_{it}})\leftarrow \delta _{\left \lfloor \lambda_{1}^{s_{it}}/k \right \rfloor}\delta _{\left \lfloor k/\lambda_{2}^{s_{it}} \right \rfloor}h_{k}^{s_{it}} $                                         \\
		17:~$\bm{ end }$ $\bm{ for }$               \\   \hline 
	\end{tabular}
\end{table}

Specifically, the design procedure in Algorithm 1 consists of four $ for  $ loops. For the first $ for $ loop, the calculation of the number of spectral bands required, which is determined by a band strcuture with $ \Lambda $, is used to minmize the value of the matrix $ H^{s}(H^{s^{'}}) $ in (4). For the second $ for $ loop, two cut-off wavelengths $ \lambda _{1}^{s_{it}} $ and $ \lambda _{2}^{s_{it}} $ obtained from the given set $ U_{ran}[argmin_{k^{'}}u_{k^{'}}] $ are exploited in order to minimize the value of the matrix $ (H_{k}^{T})H_{k^{'}} $ in (4) based on the calculation of inner product between adjacent spectral bands. Subsequently, the third $ for $ loop involves the assembling of $ h_{\lambda_{l}^{s_{it}}}^{s_{it}} $ whose entries are iid standard Gaussian random variables obtained from the set of pairs of two-cutoff wavelengths $ (\lambda _{1}^{s_{it}},\lambda _{2}^{s_{it}}) $ to minimize the inner product in the second $ for $ loop.
 
\begin{figure}[htbp]
	\centering
	
	\subfigure[]{
		\begin{minipage}[t]{0.46\linewidth}
			\centering
			\includegraphics[width=4.3cm, height=3.6cm]{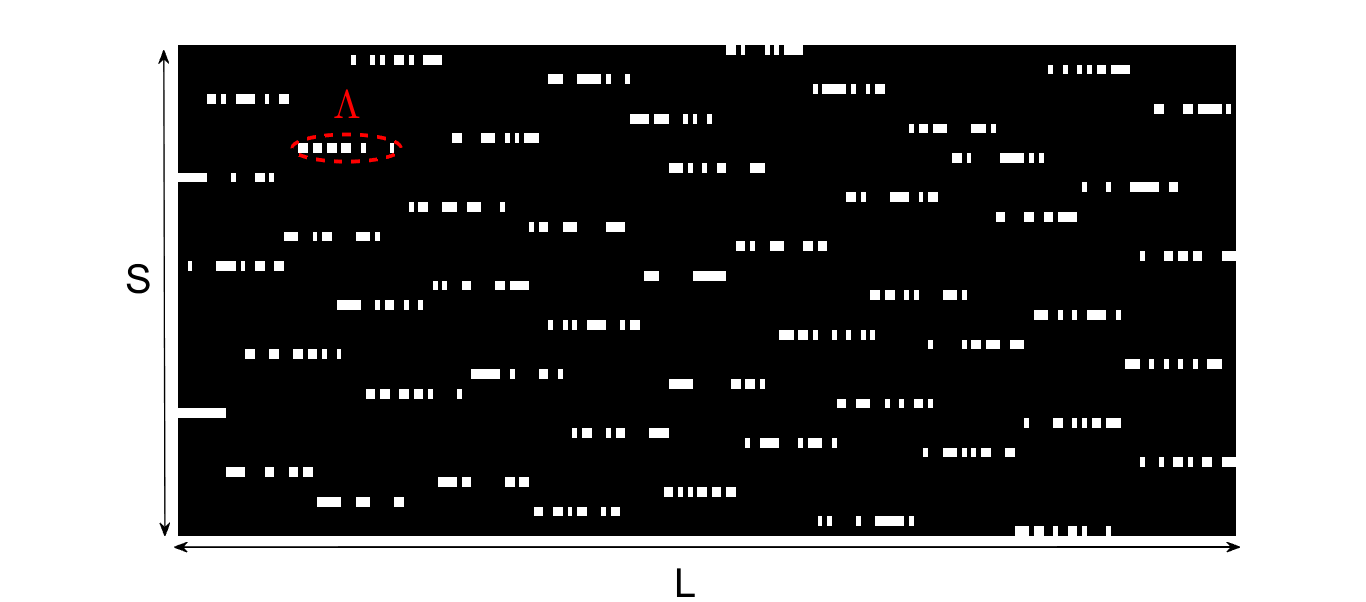}
		\end{minipage}%
	}%
	\subfigure[]{
		\begin{minipage}[t]{0.47\linewidth}
			\centering
			\includegraphics[width=4.3cm, height=3.6cm]{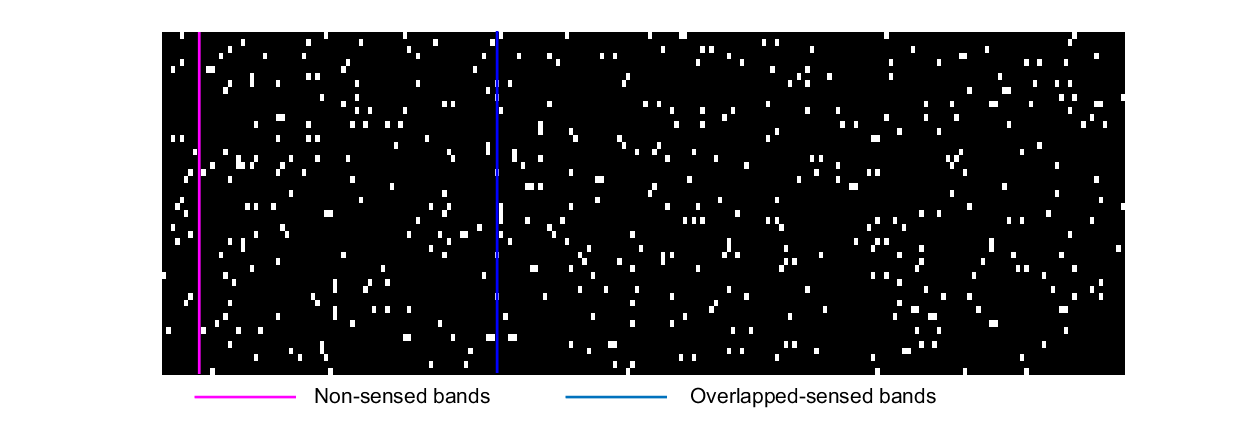}
		\end{minipage}%
	}%
	
	\centering
	\caption{Examples of the optimal coding pattern (a) and random cidong pattern (b)}
\end{figure} 

Notice in Fig. 2(a) that the block-unblock entries for the optimal coding pattern present a uniform spectral distribution providing a better sampling. Notice in Fig. 2(b) that the random coding pattern results in oversampling or unsampling of part of all spectral bands.

\section{3D-Spatial Regularized Sparse Subspace Clustering Algorithm for CSI}
\label{sec:typestyle}
We open a 3D moving window with the size of $ 3\times 3\times 3 $ at each coefficient vector and limit the difference between it and the mean of the neighboring pixels by $ \left \| c-\bar{c} \right \|_{F}^{2}< \varepsilon  $, where $ \varepsilon  $ is the restriction and $ \bar{c}\in R^{MN\times MN} $ is the mean coefficient matrix attained by rearranging the mean 3D cube $ \tilde{c}\in R^{M\times N\times MN} $ to a 2D matrix.

It is natural to introduce the 3D spatial regularization term into the SSC model framework to model sparse optimization problem in the following formulation as 
\begin{equation}\label{key}
\begin{split}
&\underset{c,g,\bar{c}}{min}\left \| c \right \|_{1}+\frac{\lambda}{2} \left \| g \right \|_{F}^{2}+\frac{\alpha}{2}\left \| c-\bar{c} \right \|_{F}^{2} \\
&s.t. ~y=yc+g, ~diag(c)=0, ~c^{T}1=1, \\
\end{split}
\end{equation}
where $ \alpha  $ is a regularization coefficient denoting the relative contribution of the spatial constraint term. The sparse optimization problem (5) can be solved with the alternating direction method of multipliers (ADMM). We then use the obtained sparse coefficient matrix $ c $ to create the weighted adjacency matrix  $ w\in R^{MN\times MN} $. The final result can then be achieved by applying the spectral clustering \cite{Ng2001On,Aldroubi2014A} to the similarity graph.

\section{SIMULATION RESULTS AND DISCUSSION}

The experiment was conducted on the Indian Pines image. This data set has $ 70\times 70 $ pixels and $ 200 $ spectral bands. This scene covers an agricultural field and contains 4 main classes: corn-no-till(2), grass(7), soybeans-no-till(10), and soybeans-minimum-till(11). There are $ 10249 $ labeled samples for this data set, with the distribution listed in Table I. The clustering is a challenging task because the spectral signatures of the land-cover classes in this area are very similar and some of the spectral curves are seriously mixed, as shown in Fig. 3(c). The false-color image and the growth truth are shown in Fig. 3(a) and (b).

\begin{figure}[htbp]
	\centering
	\subfigure[]{
		\begin{minipage}[t]{0.25\linewidth}
			\centering
			\includegraphics[width=2.5cm, height=3cm]{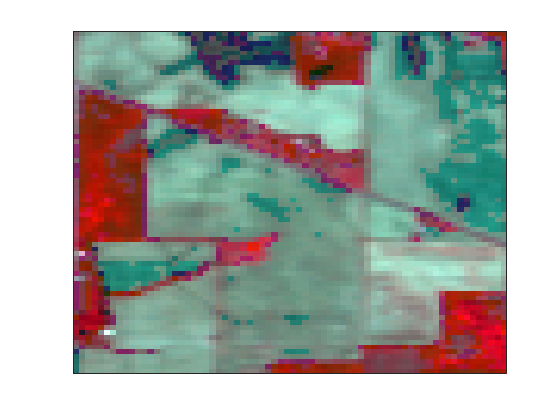}
		\end{minipage}%
	}%
	\subfigure[]{
		\begin{minipage}[t]{0.25\linewidth}
			\centering
			\includegraphics[width=2.5cm, height=3cm]{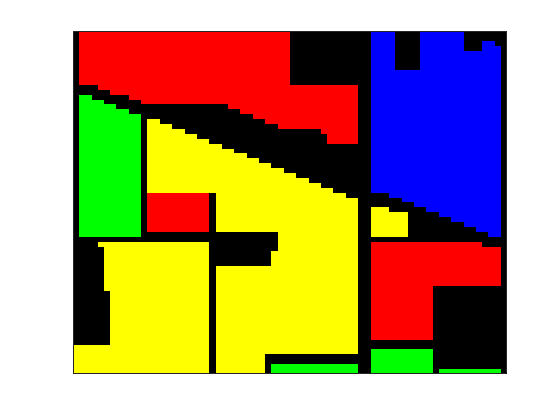}
		\end{minipage}%
	}%
	\subfigure[]{
		\begin{minipage}[t]{0.4\linewidth}
			\centering
			\includegraphics[width=3.9cm, height=3cm]{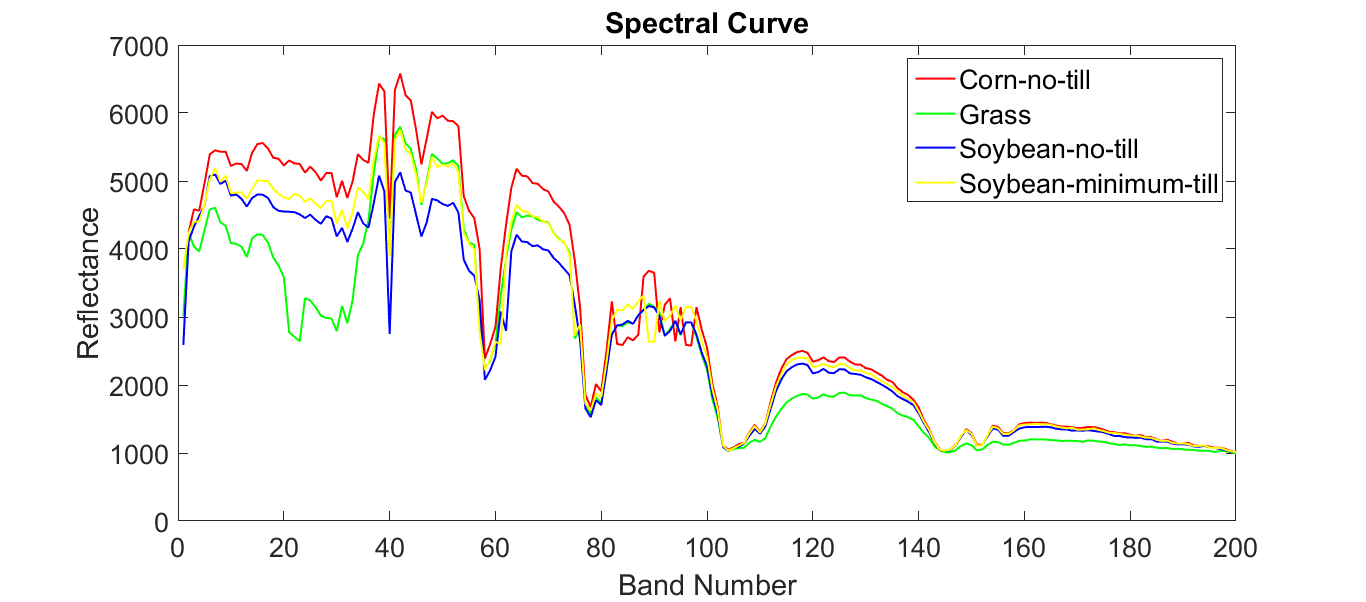}
		\end{minipage}
	}%
	\centering
	\caption{ AVIRIS indian Pines image, (a) False-color image (RGB 40, 30, 20). (b) Ground truth. (c) Spectral curves of the four land-cover classes.}
\end{figure}

\begin{table}[htbp]
	\centering
	\captionsetup{font=scriptsize}
	\caption{CLASS LABELS AND THE CORRESPONDING SAMPLE NUMBER FOR EACH CLASS OF THE INDIAN PINES DATA SET}
	\label{tab:my-table}
	\tabcolsep 0.02in
	\begin{tabular}{c|c|c|c|c|c}
		\hline
		Label & Class        & Samples & Label & Class       & Samples \\ \hline
		1     & Alfalfa      & 46      & 9     & Oats        & 20      \\
		2     & Corn-noill   & 1428    & 10    & Soy-notill  & 927     \\
		3     & Corn-mintill & 830     & 11    & Soy-mintill & 2455    \\
		4     & Corn         & 237     & 12    & Soy-clean   & 593     \\
		5     & Pasture      & 483     & 13    & Wheat       & 205     \\
		6     & Tree         & 830     & 14    & Woods       & 1265    \\
		7     & Grass        & 28      & 15    & Bidg-drives & 386     \\
		8     & Hay-window   & 478     & 16    & Stone-tower & 93      \\ \hline
		Total & \multicolumn{5}{c}{10249}                             \\ \hline
	\end{tabular}
\end{table}

The cluster map of the various clustering approaches are shown in Fig. 4(b)-(e), and the corresponding quantitative evaluation of the clustering results is provided in Talbe II, respectively, with $ 10\% $ labeled samples for training and the rest for testing. In the table, the optimal value of each row is shown in bold and the second best results are underlined. From Fig. 4 and Table II, it can be clearly observed that the Optimal-codes-3D-SRSSC and Full-data-3D-SRSSC obtain a better accuracy by making use of the spatial neighborhood information. Also, compared with Random-codes-3D-SRSSC, Optimal-codes-3D-SRSSC performs better, obtaining a higher accuracy, which demonstrates the optimal codes approximately preserve the similarities among spectral pixels. Finally, the proposed approach obtains clustering results comparable to the classification of the full data by the Full-data-3D-SRSSC and Full-data-SSC. Nevertheless, the proposed approach reduces the computational time by $ 83.65\% $ and $ 83.23\% $ compared to Full-data-3D-SRSSC and Full-data-SSC.

\begin{figure}[htbp]
	\centering
	
	\subfigure[]{
		\begin{minipage}[t]{0.3\linewidth}
			\centering
			\includegraphics[width=3cm, height=3cm]{figure/figure17}
		\end{minipage}%
	}%
	\subfigure[]{
		\begin{minipage}[t]{0.3\linewidth}
			\centering
			\includegraphics[width=3cm, height=3cm]{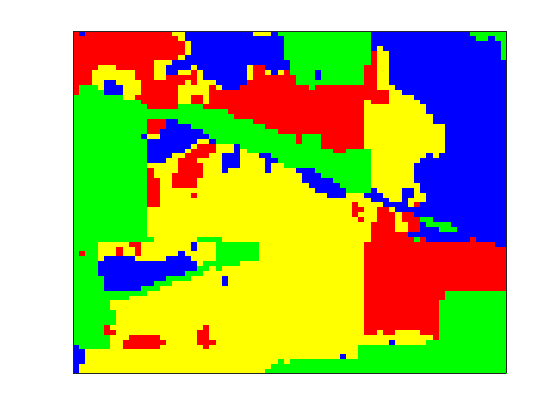}
		\end{minipage}%
	}%
	\subfigure[]{
		\begin{minipage}[t]{0.3\linewidth}
			\centering
			\includegraphics[width=3cm, height=3cm]{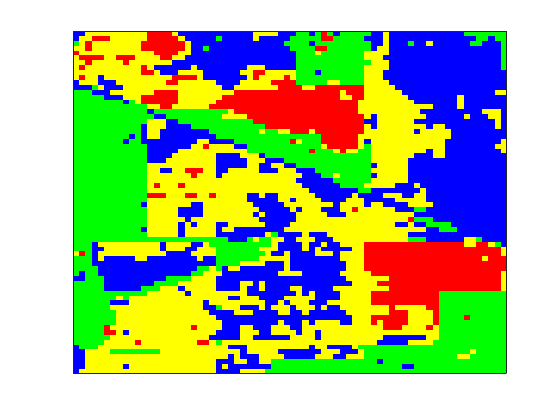}
		\end{minipage}%
	}%
	
	\subfigure{
		\begin{minipage}[t]{0.3\linewidth}
			\centering
			\includegraphics[width=2.9cm, height=2.6cm]{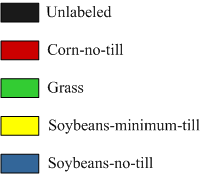}
		\end{minipage}
	}%
	\subfigure[]{
		\begin{minipage}[t]{0.3\linewidth}
			\centering
			\includegraphics[width=3cm, height=3cm]{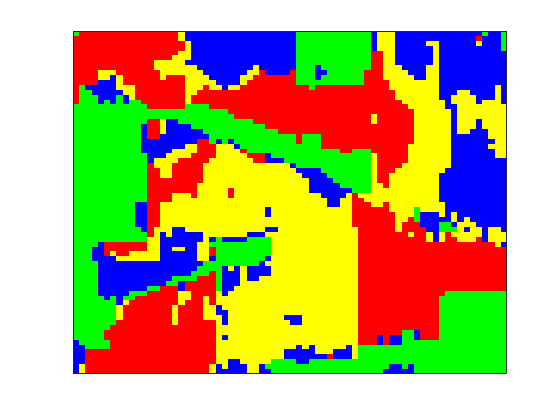}
		\end{minipage}
	}%
	\subfigure[]{
		\begin{minipage}[t]{0.3\linewidth}
			\centering
			\includegraphics[width=3cm, height=3cm]{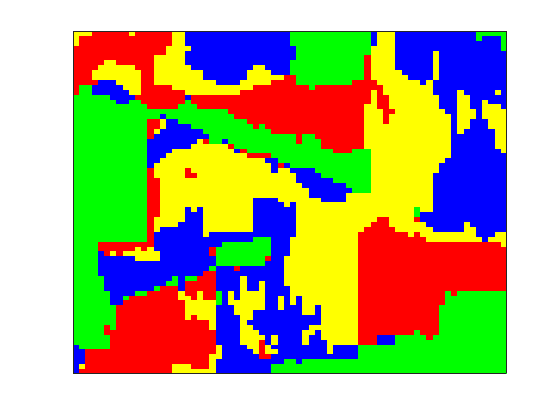}
		\end{minipage}
	}%
	
	\centering
	\caption{ Cluster maps of different approaches with the Indian Pines image: (a) Ground truth. (b) Full-data-3D-SRSSC, (c) Full-data-SSC, (d) Optimal-codes-3D-SRSSC, (e) Random-codes-3D-SRSSC.}
\end{figure}         

\begin{table}[htbp]
	\centering
	\captionsetup{font=scriptsize}
	\caption{QUANTITATIVE EVALUATION OF THE DIFFERENT CLUSTERING APPAROAHCES FOR THE INDIAN PINES IMAGE}
	\label{tab:my-table}
	\tabcolsep 0.02in
	\begin{tabular}{ccccc}
		\hline
		Class        & Random & Optimal & Full--SSC & Full-3D-SRSSC \\ \hline
		2            & $ \mathbf{73.13} $        & $ \underline{71.45} $         & 48.96         & 66.77              \\
		7            & 95.25        & $ \mathbf{100} $           & $ \underline{98.60} $         & $ \mathbf{100} $                \\
		10           & 52.58        & $ \mathbf{89.36} $         & $ \underline{70.63} $         & 69.54              \\
		11           & 55.29        & $ \underline{63.52} $         & 59.23         & $ \mathbf{80.05} $              \\ \hline
		OA           & 63.83        & $ \underline{74.15} $         & 62.62         & $ \mathbf{76.16} $              \\
		AA           & 69.14        & $ \mathbf{81.57} $         & 69.35         & $ \underline{79.09} $              \\
		Kappa        & 49.26        & $ \underline{63.78} $         & 47.58         & $ \mathbf{65.89} $              \\ \hline
		Time {[}s{]} & 46.40        & 30.30         & 283.88        & 179.13             \\ \hline
	\end{tabular}
\end{table}

\bibliographystyle{IEEEtran}
\bibliography{reference1}

\end{document}